\documentclass[iop,apj,numberedappendix,appendixfloats]{emulateapj}

\usepackage{amsmath}
\usepackage{multirow}
\usepackage{graphicx,float}
\usepackage{gensymb}

\bibpunct[, ]{(}{)}{;}{a}{}{,}

\begin{document}

\title{Probing Galactic Structure with the Spatial Correlation Function of SEGUE G-dwarf Stars}

\shorttitle{The Clustering of SEGUE G-dwarf Stars}

\author{
  Qingqing~Mao\altaffilmark{1},
  Andreas~A.~Berlind\altaffilmark{1,2},
  Kelly~Holley-Bockelmann\altaffilmark{1},
  Katharine~J.~Schlesinger\altaffilmark{3},
  Jennifer~A.~Johnson\altaffilmark{4},
  Constance~M.~Rockosi\altaffilmark{5},
  Timothy~C.~Beers\altaffilmark{6},
  Donald~P.~Schneider\altaffilmark{7}
  Kaike~Pan\altaffilmark{8}
  Dmitry~Bizyaev\altaffilmark{8,9}
  Elena~Malanushenko\altaffilmark{8}
}
\altaffiltext{1}{Department of Physics and Astronomy, Vanderbilt University, Nashville, TN 37235, USA}
\altaffiltext{2}{a.berlind@vanderbilt.edu}
\altaffiltext{3}{Research School of Astronomy and Astrophysics, The Australian National University, Weston, ACT 2611, Australia}
\altaffiltext{4}{Department of Astronomy, The Ohio State University, Columbus, OH 43210, USA}
\altaffiltext{5}{UCO/Lick Observatory, Department of Astronomy and Astrophysics, University of California, Santa Cruz, CA 95064, USA}
\altaffiltext{6}{Department of Physics and JINA Center for Evolution of the Elements, University of Notre Dame, Notre Dame, IN  46556, USA}
\altaffiltext{7}{Institute for Gravitation and the Cosmos, Department of Astronomy and Astrophysics, The Pennsylvania State University, University Park, PA 16802, USA}
\altaffiltext{8}{Apache Point Observatory and New Mexico State University, P.O. Box 59, Sunspot, NM, 88349-0059, USA}
\altaffiltext{9}{Sternberg Astronomical Institute, Moscow State University, Moscow, Russia}

\begin{abstract}
\label{abstract}
We measure the two-point correlation function of G-dwarf stars within $1-3$~kpc of the Sun in multiple lines-of-sight using the Schlesinger~et~al. G-dwarf sample from the SDSS SEGUE survey. The shapes of the correlation functions along individual SEGUE lines-of-sight depend sensitively on both the stellar-density gradients and the survey geometry. We fit smooth disk galaxy models to our SEGUE clustering measurements, and obtain strong constraints on the thin- and thick-disk components of the Milky Way. Specifically, we constrain the values of the thin- and thick-disk scale heights with 3\% and 2\% precision, respectively, and the values of the thin- and thick-disk scale lengths with 20\% and 8\% precision, respectively. Moreover, we find that a two-disk model is unable to fully explain our clustering measurements, which exhibit an excess of clustering at small scales ($\lesssim 50$~pc). This suggests the presence of small-scale substructure in the disk system of the Milky Way.
\end{abstract}


\section{Introduction}
\label{s:introduction}
The Milky Way provides a unique laboratory for studying the structure of a galaxy in detail, by allowing us to measure and analyze the properties of large samples of individual stars (see reviews by \citealt{Ivezic:2012} and \citealt{Rix:2013}). Recent surveys, such as the Sloan Digital Sky Survey (SDSS I-III; \citealt{York:2000, Eisenstein:2011}), the Two-Micron All Sky Survey (2MASS; \citealt{Skrutskie:2006}), the Radial Velocity Experiment (RAVE; \citealt{Kordopatis:2013}), and others have placed strong constraints on the smooth components of the Milky Way \citep[e.g.,][]{Carollo:2010, Bovy:2012}, and have discovered significant spatial substructure in the Milky Way, such as stellar streams \citep[e.g.,][]{Belokurov:2006} and stellar overdensities \citep[e.g.,][]{Juric:2008}.  Investigating the structure of the Milky Way provides clues about galaxy formation and evolution that cannot be extracted from observations of distant galaxies. 

The Sloan Extension for Galactic Understanding and Exploration (SEGUE; \citealt{Yanny:2009}) is a spectroscopic sub-survey of the SDSS that focused on Galactic science. SEGUE data provides the largest spectroscopic sample of Galactic stars currently available, and covers a more extensive volume of the Milky Way than previous studies, probing from the local disk all the way to the outer stellar halo. The full SEGUE survey provides an unprecedented opportunity to investigate the structure of the Milky Way \citep[e.g.,][]{Carollo:2010,deJong:2010,Cheng:2012}.

The spatial two-point correlation function is one of the simplest and most effective statistical tools for studying clustering in general, and it is widely used in studies of the large-scale structure of the Universe (\citealt{Peebles:1973}; see \citealt{Anderson:2014} for a recent example). However, it has rarely been used in Galactic structure studies, mainly due to the lack of large and homogeneous spectroscopic stellar samples. There have only been a few applications of the correlation function applied to Galactic halo stars, especially giants and blue horizontal-branch (BHB) stars, but the sample sizes were limited. \citet{Doinidis:1989} analyzed over 4,400 BHB stars, and found an excess correlation with separations $r \leq 25$ pc. \citet{Starkenburg:2009} developed a phase-space correlation function, and applied it to 101 giants in the Spaghetti project \citep{Morrison:2000} to search for substructures in the halo. The phase-space correlation function has also been applied to various BHB samples to quantify the amount of spatial and kinematic substructure in the Milky Way's stellar halo \citep{DePropris:2010, Xue:2011, Cooper:2011}. In addition to these spatial studies, the angular two-point correlation function has been used to study the stellar cluster distribution \citep{Lopez-Corredoira:1998} and to search for wide binaries (see \citealt{Longhitano:2010} as an example). 

With the advent of large stellar samples provided by the SEGUE survey, it is time to explore Galactic structure by applying the correlation function to stars. In this article, we measure the full 3-D spatial two-point correlation function of the SEGUE G-dwarf sample, which is the largest stellar category in the survey. In \S\ref{s:sample} we describe the basics of the SEGUE survey and the G-dwarf sample we use. In \S\ref{s:measurement} we present our correlation function measurements and build intuition about its shape by investigating how it depends on the underlying stellar-density gradient and survey geometry. In \S\ref{s:mcmc} we fit a smooth Galactic model to our measurements and in \S\ref{s:substructure} we study residuals with respect to this model. Finally, we summarize our results and discuss possible future work in \S\ref{s:summary}.

\section{The SEGUE G-dwarf Sample}
\label{s:sample}

\begin{figure}
	\includegraphics[scale=0.35]{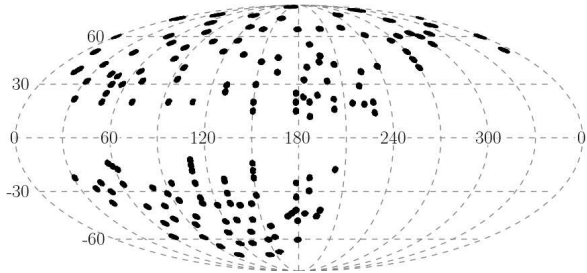}
	\caption{Sky map of the 152 SEGUE fields used in this study, shown in a Mollweide projection in Galactic coordinates. Each point indicates the location of a single pencil-beam volume that is probed by a SEGUE spectroscopic plate covering 7 deg$^2$ on the sky.}
	\label{fig:geometry_mollweide}
\end{figure}

\begin{figure}
	\includegraphics[scale=0.32]{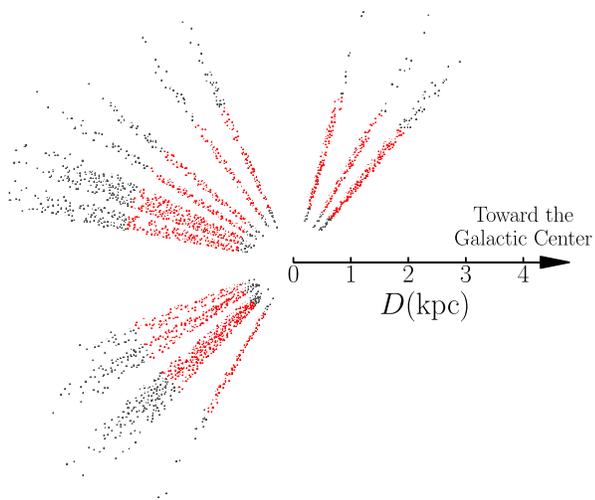}
	\caption{A selection of SEGUE pencil-beam fields in a slice perpendicular to the Galactic plane, including the Galactic center. Specifically, the slice shows fields with Galactic longitudes within ten degrees of $0\degree$ or $180\degree$ Galactic longitude. Each dot shows the location of a SEGUE G-dwarf, with red points indicating stars with distances between $1-3$~kpc, which are used in this study.}
	\label{fig:geometry_pencils}
\end{figure}

The SEGUE survey makes use of the dedicated SDSS telescope \citep{Gunn:2006} and multi-object spectrograph \citep{Smee:2013}. SEGUE combines the extensive and uniform photometry from the SDSS with medium-resolution ($R \sim 1800$) spectroscopy over a broad spectral range ($3800-9200\AA$) for $\sim$ 240,000 stars spanning a range of spectral types. SEGUE was designed to sample Galactic structure at a variety of distances in $\sim$ 200 pencil-beam' volumes spread over the sky available from Apache Point. Each pencil beam corresponds to a single SDSS spectroscopic plate covering a circular region of 7 square degrees and probes a selection of stars in that line-of-sight with up to 640 spectroscopic fibers \citep{Yanny:2009}. Figure~\ref{fig:geometry_mollweide} displays the sky positions of the pencil beams included in this study using a Mollweide projection in Galactic coordinates. Figure~\ref{fig:geometry_pencils} presents an edge-on view of the pencil beams with Galactic longitudes near the Galactic center and the Galactic anticenter.

The G-dwarf sample represents SEGUE's largest single homogeneous stellar spectral target category. The SEGUE G dwarfs are defined as having magnitudes and colors in the range $14.0 < r_0 < 20.2$ and $0.48 < (g-r)_0 < 0.55$, where $g_0$ and $r_0$ are the extinction-corrected $g$- and $r$-band magnitudes (the extinction correction uses the \citealt{Schlegel:1998} dust map). This simple target selection makes the selection biases relatively straightforward to understand \citep{Yanny:2009}. Here we use the G-dwarf catalog with distances and weights derived by \citet{Schlesinger:2012}. Distances are estimated with an isochrone-matching technique that is accurate to $\sim 8\%$ for metal-poor and $\sim 18\%$ for metal-rich stars \citep{An:2009}.

\begin{figure*}
	\includegraphics[scale=0.8]{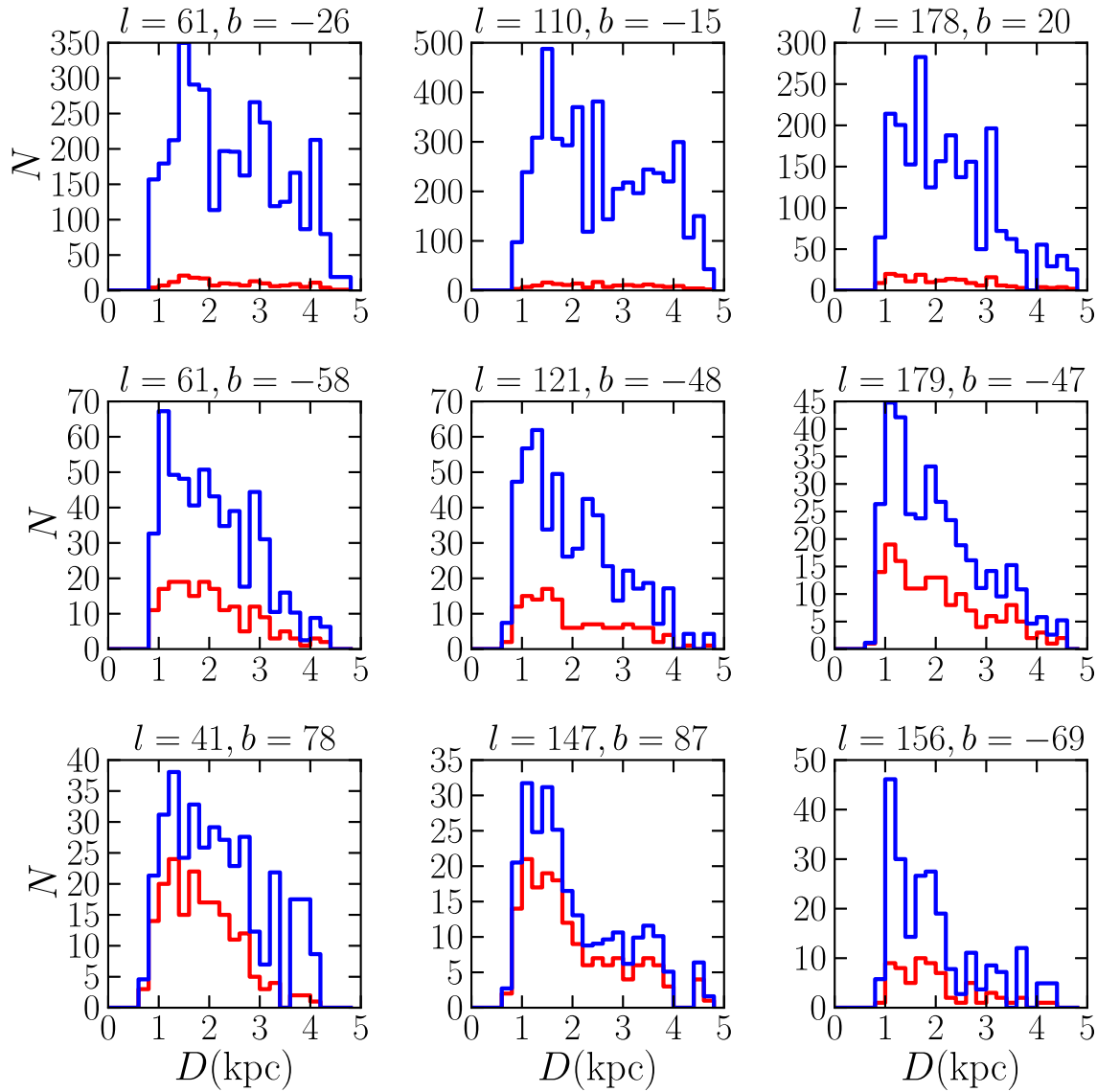}
	\caption{Distribution of G-dwarf stars with distance, along a selection of nine SEGUE lines-of-sight. Each panel shows a particular SEGUE field, with panels arranged so that, going from top to bottom, fields point farther away from the Galactic plane in latitude, and, going from left to right, fields point farther away from the Galactic center in longitude. The Galactic coordinates of each field are listed at the top of each panel. The unweighted distributions are shown in red, while the weighted distributions, which are corrected for incompleteness, are shown in blue (see \citealt{Schlesinger:2012} for more details on the weighting scheme employed).}
	\label{fig:distance_distr}
\end{figure*}

We also apply the target-type weights and the $r$-magnitude weights in the catalog described by \citet{Schlesinger:2012} to correct for various selection biases. SEGUE categories often focus on specific ranges in parameter space, and targets that fulfill multiple target-type criteria have multiple opportunities to be assigned a spectroscopic fiber. This approach leads to a slightly biased G-dwarf selection, which can be corrected for by the target-type weights. SEGUE assigns roughly the same number, $\sim300$, of spectroscopic fibers to G-dwarf targets on each plug-plate, but this is far less than the actual number of available G-dwarfs, which also varies from field-to-field. As the stellar number density changes over the SEGUE footprint, we must use $r$-magnitude weights to correct for this variable sampling, in order to better represent the true underlying stellar distribution in the Milky Way. For more details about the survey completeness and weights, we refer readers to \citet[][$\S 4.7$]{Schlesinger:2012}. Figure~\ref{fig:distance_distr} presents the distribution of G-dwarf stars with distance for a selection of nine SEGUE fields of varying Galactic latitude and longitude. The figure shows both the raw and weighted stellar distributions. Although the different lines-of-sight contain similar numbers of G-dwarf stars (as seen from the unweighted distributions), it is clear that there are large differences in the weighted distributions. Fields near the Galactic disk and the Galactic center have larger $r$-weights to account for the denser stellar distributions in those directions.

To achieve a sufficiently high number density of stars throughout our sample volume, and to avoid unrealistically large weights at the near and far ends of the pencil beams, we restrict the sample to stars with distances from $1-3$~kpc, and ignore pencil beams containing less than 50 G dwarfs. These selection criteria produce a sample of 18,067 G dwarfs in 152 pencil beams that we use in our analysis. 

\section{Two-point Correlation Function Measurements}
\label{s:measurement}

In galaxy surveys, a common method to estimate the correlation function of a given sample is to construct a denser and uniform random sample with the same survey geometry, and then, in each distance separation bin $[r, r + \Delta r]$, count the number of pairs in both the survey data and the random sample. The correlation function can then be estimated by the so-called natural estimator, 
\begin{equation}
  \label{eq:xi}
  \xi(r) = \frac{DD(r)}{RR(r)} - 1, 
\end{equation}
where $DD$ are the weighted and normalized pair counts of objects found in each separation bin, and $RR$ are the normalized pair counts of random points. The two terms are normalized by dividing by the square of the total number of data and random points, respectively. When estimating the correlation function of galaxies, it makes sense to use a uniformly distributed random sample because the universe is intrinsically homogeneous and isotropic on large scales. However, this is not the case for stars within the Galaxy, which are distributed in disk and halo structures that exhibit strong global density gradients.

If we know the global spatial-density distribution of stars in the Galaxy, we can construct a substitute for the random sample that instead follows the same global distribution as the stars. The measured correlation function will then mostly cancel on all scales, and reveal whatever excess clustering remains. If we do not fully know the underlying density distribution of stars, we can still compare the observed data to a uniform random sample, but then the measured correlation function will have a shape that encodes this information. The pencil-beam survey geometry can also add complications. The interplay between the survey geometry and the non-uniform density distribution of stars can create additional signals in the correlation function. 

\begin{figure}
	\includegraphics[scale=0.3]{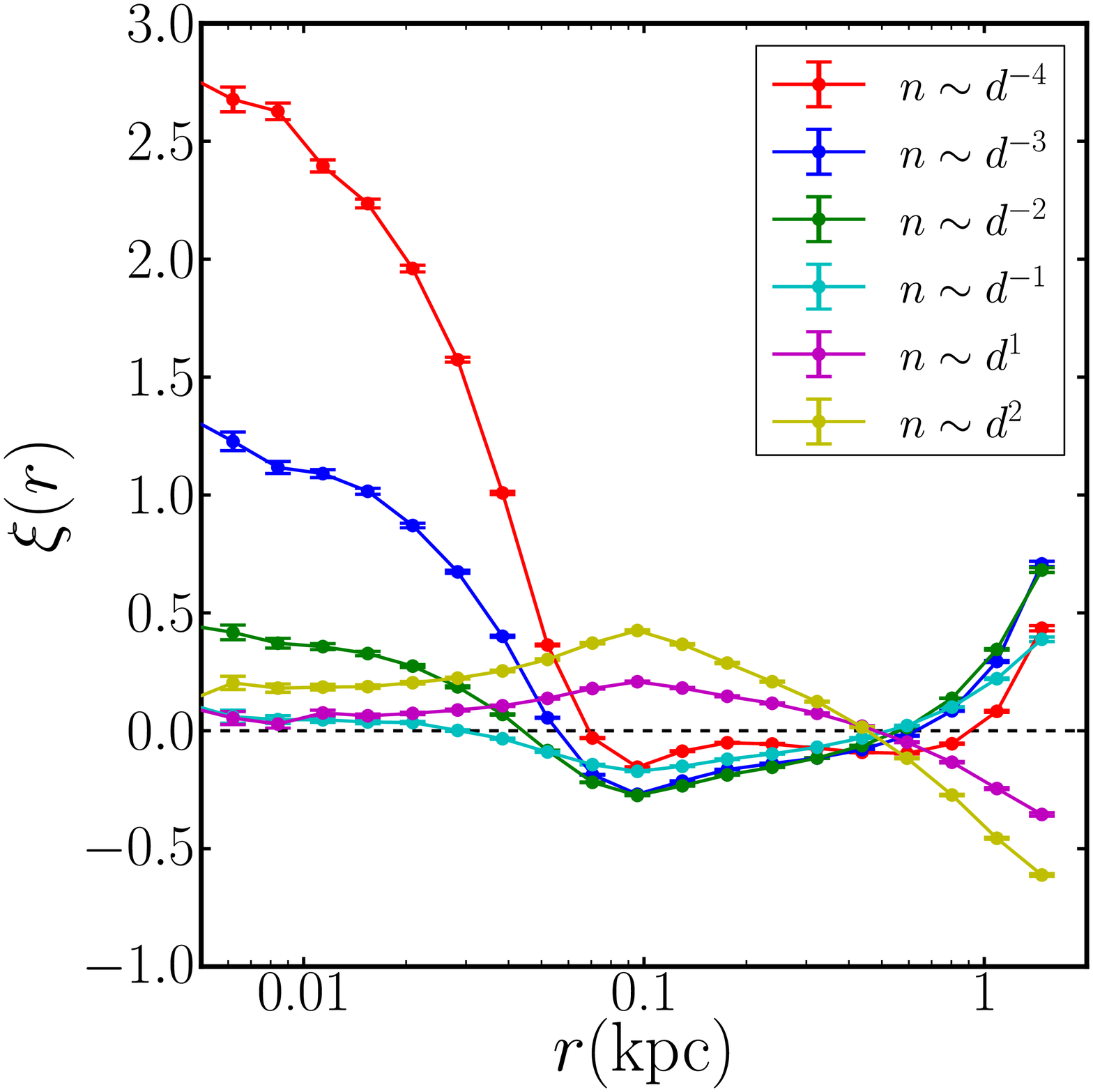}
	\caption{Dependence of the correlation function on the underlying density gradient. The correlation functions are computed for mock star samples that all have the same pencil-beam geometry as one of our SEGUE lines-of-sight, but are designed to have different power-law stellar density profiles, as listed in the panel. Each curve is the average over 1000 mock samples containing 1000 stars each; the error bars show the uncertainty in the mean as estimated from the standard deviation among the 1000 mocks. The correlation function has a complex shape, and is highly sensitive to the stellar-density gradient, especially on small scales.}
	\label{fig:test_gradient}
\end{figure}

Before computing the correlation function of the SEGUE stars, we first investigate how stellar-density gradients and the pencil-beam survey geometry can affect the shape of the correlation function in general, by creating different mock star samples and measuring their correlation function. First, we set the mock survey geometry to be the same as that in one of our SEGUE lines-of-sight, i.e., a pencil beam with an angular diameter of $3\degree$ and distances between $1-3$~kpc. We generate mock star samples within this geometry, each containing 1000 mock stars, using different power-law density profiles. Specifically, the density gradients we adopt are $n\sim d^{-4}$, $d^{-3}$,  $d^{-2}$,  $d^{-1}$,  $d^{1}$,  and $d^{2}$, where $n$ is the number density of stars and $d$ is the distance from the observer. We then construct a uniformly distributed random sample with ten times the number density, and calculate the correlation function using equation~\ref{eq:xi} for each density profile. Finally, we repeat these steps 1000 times, using independent realizations of the mock samples, and average the results to reduce the noise. Note that these mock samples with power-law density gradients are not meant to represent realistic Galactic models, but serve the purpose of building intuition on how density gradients can affect the derived correlation function. We study realistic Galactic models in \S~\ref{s:mcmc}.

Figure~\ref{fig:test_gradient} shows the resulting correlation functions for our adopted density gradients. The overall shape of the correlation function is quite complex, and is very sensitive to the density profile. On small scales ($\lesssim50$~pc) the correlation function is always boosted, regardless of whether the underlying density gradient is positive or negative. However, on larger scales ($\sim50-500$~pc), the clustering is depressed or boosted depending on whether the density gradient is negative or positive, respectively. Finally, on even larger scales ($\gtrsim500$~pc) the sign of this dependence flips. There is an interesting feature at $r\sim0.1$~kpc, where the correlation function is a minimum or maximum for negative and positive density gradients, respectively. This is approximately equal to the diameter of the pencil beam volume at its halfway point along the line-of-sight.

\begin{figure}
	\includegraphics[scale=0.3]{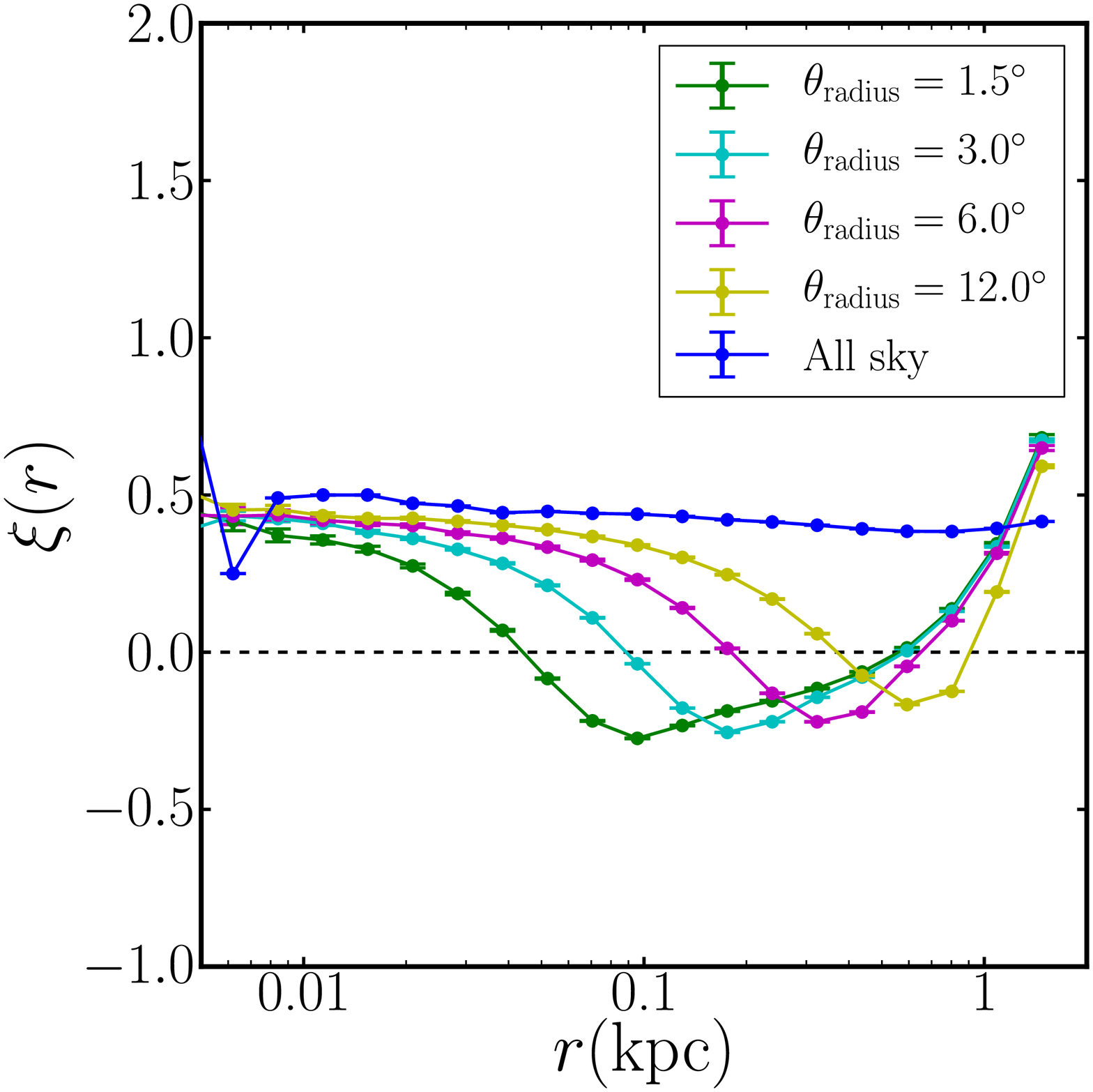}
	\caption{Dependence of the correlation function on survey geometry. The correlation functions are computed for mock star samples that all have the same stellar-density profile of $n\sim d^{-2}$, but occupy different sample geometries. All sample geometries range in distance from $1-3$~kpc, but their angular extent on the sky varies from a circle of radius $\theta_\mathrm{radius} = 1.5\degree$, all the way up to the full sky, as listed in the panel. As in Fig.~\ref{fig:test_gradient}, points and errors are estimated from 1000 mock samples. The correlation function is sensitive to the survey geometry, featuring a minimum at a scale approximately equal to the diameter of the pencil-beam volume at its halfway point along the line-of-sight.}
	\label{fig:test_geometry}
\end{figure}

Next, we investigate how the survey geometry can affect the correlation function. We set the underlying density profile of our mock star samples to be $n\sim d^{-2}$, and vary the sample geometry. The geometries we test all range in distance from $1-3$~kpc, as in our SEGUE sample, but their angular size on the sky varies from the pencil beam of radius $\theta_\mathrm{radius} = 1.5\degree$ (as in SEGUE), to larger beams of radius $3\degree$, $6\degree$, $12\degree$, as well as a full-sky geometry. As before, we generate 1000 independent mock samples for each geometry, and calculate their average correlation function using a uniform and dense random sample. Figure~\ref{fig:test_geometry} reveals a fairly simple dependence of the correlation function on survey geometry. On scales that are much smaller than the width of the pencil beam, the correlation function is unaffected by the survey geometry, as might be expected. However, the feature in the correlation function that occurs at 0.1~kpc for the SEGUE geometry shifts to progressively larger scales as the width of the pencil beam grows. In fact, the scale of the feature is always approximately equal to the diameter of the pencil-beam volume at its halfway point along the line-of-sight.

These tests demonstrate that the correlation function of stars will depend sensitively on both the underlying density gradients and the survey geometry. The resulting correlation function has a peculiar shape that is quite different from the power-law shape we are accustomed to seeing for galaxy surveys. The strong dependence on the underlying density gradients suggests that the correlation function of stars could have strong constraining power on Galactic structure models. This is especially true at the smallest scales and when density gradients are steep, since this is where the correlation function is most sensitive to variations in the underlying density distribution. The explanation for this is fairly straightforward. At the smallest scales we probe ($\lesssim 10$pc), the mean separation between stars is much larger, and so there would not be many pairs if the stars were randomly distributed. If, however, there is a steep enough density gradient, the stars are redistributed so that they become sufficiently dense at either the near or far end of the survey volume (depending on whether the gradient is negative or positive), thus leading to several small-scale pairs.

\begin{figure}
	\includegraphics[scale=0.3]{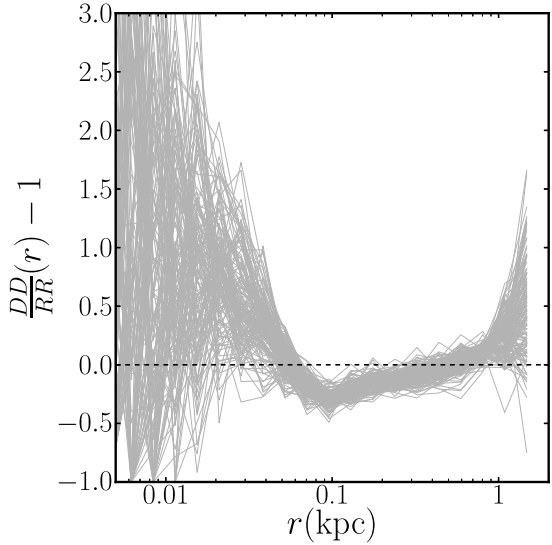}
	\caption{The two-point correlation functions of SEGUE G-dwarf stars. Each gray line is measured from one of 152 individual SEGUE lines of sight. The shapes of the correlation functions are similar to those for the negative gradients shown in Fig.~\ref{fig:test_gradient}.}
	\label{fig:correlation_data}
\end{figure}

To measure the correlation function of SEGUE G-dwarf stars, we first construct a random sample with the same pencil-beam geometry as our sample, and containing uniformly distributed points with 100 times higher number density than the SEGUE data. We then calculate the correlation function of each SEGUE line-of-sight independently, i.e., we only count pairs of stars that reside in the same SEGUE field. Figure~\ref{fig:correlation_data} shows the result, in which each gray line is the correlation function of an individual SEGUE pencil beam. The measured correlation functions have the same peculiar shape seen in the mock tests in Figure~\ref{fig:test_gradient}. In particular, they are similar to the cases of negative density gradients, which makes perfect sense, since all SEGUE lines of sight move out of the Galactic disk.

The distances to the SEGUE stars are not known perfectly, but rather contain, on average, 12\% uncertainties. It is thus important to determine how much these errors can affect the correlation function measurements. We test this issue by adding $12\%$ Gaussian-distributed distance errors to our mock samples, and then recalculating the correlation functions. These tests demonstrate that $12\%$ distance uncertainties have a negligible effect on the correlation function.

\section{Fitting A Smooth Galactic Model}
\label{s:mcmc}

Since the two-point correlation function of G dwarfs is highly sensitive to stellar-density gradients, it can serve as a tool to probe the smooth density structure of the Milky Way. We approach this by replacing the uniform random sample in equation \ref{eq:xi} with a mock sample generated from a Milky Way model, 
\begin{equation}
  \label{eq:xip}
  \xi'(r) = \frac{DD(r)}{MM(r)} - 1,  
\end{equation}
where $MM$ are the normalized pair counts from our Milky Way model. If the model we choose truly represents the underlying stellar distribution and has the same geometry as the data, then $\xi'(r)$ should cancel on all scales and along all lines-of-sight. By searching the parameter space of a given model, we can thus place constraints on the model parameters, and determine to what extent the model can explain the observed stellar clustering.

As a proof of concept, we choose a standard thin- + thick-disk model with two exponential disk components and five parameters, 
\begin{equation}
\label{eq:model}
\begin{aligned}
n(R, Z) \propto \textmd{sech}^2\left(\frac{Z}{2 Z_{0,\mathrm{thin}}}\right)\textmd{exp}\left(-\frac{R}{R_{0,\mathrm{thin}}}\right) \\
+ a\ \textmd{sech}^2\left(\frac{Z}{2Z_{0,\mathrm{thick}}}\right)\textmd{exp}\left(-\frac{R}{R_{0,\mathrm{thick}}}\right), 
\end{aligned}
\end{equation}
where $Z_{0,\mathrm{thin}}$, $Z_{0,\mathrm{thick}}$, $R_{0,\mathrm{thin}}$, $R_{0,\mathrm{thick}}$ are scale heights and scale lengths of the thin disk and the thick disk, respectively, and the fifth parameter is the ratio of the normalization factors of the thick and the thin disk, $a = n_{0,\mathrm{thick}} / n_{0,\mathrm{thin}}$. In a recent study, \citet{Bovy:2012a} reported that when one separates disk populations by their chemical signatures, there is a continuous range of disk thicknesses, and there is no distinct thick disk component. Since we do not apply any additional color or metallicity cuts in the sample, for simplicity we stick to the traditional bi-modal disk model. Our model does not include a bulge or halo component because, in the restricted range of distances we probe ($1-3$~kpc), these components should contribute a negligible number of stars to our sample.

We employ a Markov-chain Monte Carlo (MCMC) method to identify the region in parameter space where $\xi'(r)$ is consistent with zero, i.e., to find the parameters that best fit the SEGUE clustering data. At every MCMC step, we need to have a mock catalog from our model that is generated from a given set of parameter values and has the same SEGUE survey geometry (all lines-of-sight). Moreover, the mock catalog should be substantially denser than the SEGUE data, so that the errors in $MM$ are much smaller than the errors in $DD$. Generating new dense mock samples and finding pairs at each step of the chain can be computationally expensive. Instead, we adopt a strategy that is both accurate and more efficient. We first generate a single dense and uniformly distributed random sample with the SEGUE geometry (all lines-of-sight) and identify all the pairs of points in bins of separation. At each step in the chain we assign a new weight, $w_i$, to each random point according to equation \ref{eq:model}. We then calculate $MM(r)$ by summing the product $w_iw_j$ over all pairs with separation $r$. Finally, we normalize $MM$ by the sum of $w_iw_j$ over all pairs and all scales. When normalizing, the absolute normalization of $n(R,Z)$ cancels and is thus irrelevant.

\begin{figure}
	\includegraphics[scale=0.4]{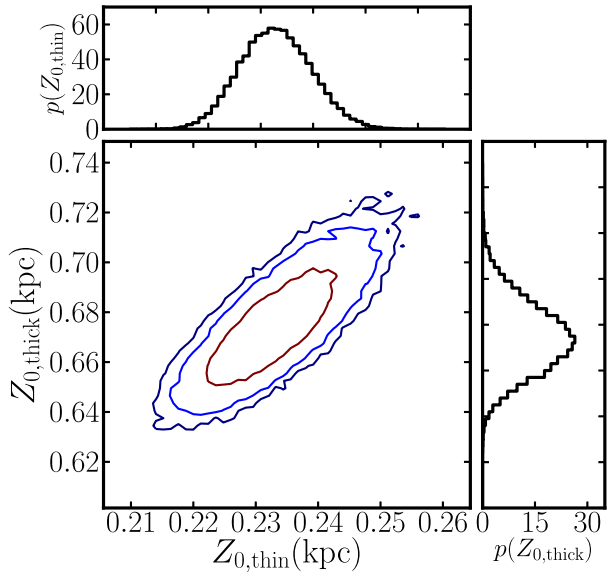}
	\caption{The distribution of scale heights for the thin and the thick disk from the MCMC chain. The main panel shows 1-, 2-, and 3$\sigma$ likelihood contours for the joint probability distribution of both scale heights, while the smaller panels on top and to the right show the individual probability distributions of each scale height, marginalized over all other parameters. The 1-$\sigma$ statistical precision of these constraints is 3\% and 2\% for the thin- and thick-disk scale heights, respectively.}
	\label{fig:mcmc_result_z}
\end{figure}

In each of our 152 pencil-beam volumes, we calculate $\xi'(r)$ in 12 logarithmic bins ranging from 5~pc to 2~kpc. Excluding any bins that have zero pair counts in $DD$, we have 1,777 individual measurements of $\xi'_i(r)$. We estimate the total $\chi^2$ using
\begin{equation}
  \chi^2 = \sum\limits_{i,r} \frac{\xi'^2_i(r)}{\sigma_i^2(r)},
\end{equation}
which sums over all scales and all pencil beams. We use jackknife resampling to estimate the uncertainties of pair counting in both the data and the model. The final uncertainty, $\sigma_i^2(r)$, is a combination of the uncertainty in the data and the uncertainty in our model, although the pair counting in our model always has much smaller uncertainties than the data because it has a much higher number density. We treat all $\xi'_i(r)$ as independent measurements and ignore the covariances. We will investigate the covariances in a future study. 

\begin{figure}
	\includegraphics[scale=0.4]{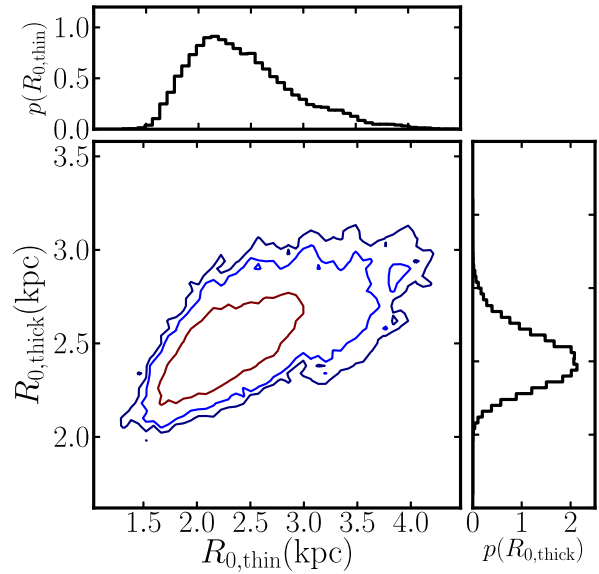}
	\caption{The distribution of scale lengths of the thin and the thick disk from the MCMC chain. All features are similar to those in Fig.~\ref{fig:mcmc_result_z}. The 1-$\sigma$ statistical precision of these constraints is 20\% and 8\% for the thin- and thick-disk scale lengths, respectively.}
	\label{fig:mcmc_result_r}
\end{figure}

Figures~\ref{fig:mcmc_result_z} and~\ref{fig:mcmc_result_r} show the scale-height and scale-length distributions from our MCMC chains. After marginalizing over all other parameters, we obtain a thin-disk scale height of $233\pm7$~pc and scale length of $2.34\pm0.48$~kpc, and a thick-disk scale height of $674\pm16$~pc and scale length of $2.51\pm0.19$~kpc. While these numbers are in the same broad range as other recent measurements using SEGUE or SDSS data \citep[e.g.,][]{Juric:2008,Carollo:2010,deJong:2010,Bensby:2011,Cheng:2012,Bovy:2012}, they are not in statistical agreement with most of these studies. Unfortunately, it is difficult to directly compare our results to other studies because the star samples differ significantly in most cases (i.e., different types of stars or different metallicity or color cuts). For example, our thick-disk scale height and length are significantly lower than those measured by \citet{Juric:2008}, but that study used M stars from the SDSS. Our thick-disk scale length is significantly higher than the one measured by \citet{Cheng:2012}, who also used SEGUE data, but that study focused on $\alpha$-enhanced stars. Our thin-disk scale height and length are somewhat smaller than those measured by \citet{Bovy:2012}, but in that study 'thin' and 'thick' disks refer to single disk fits to either $\alpha$-young or $\alpha$-old G-dwarf subsamples, respectively. It would be interesting to repeat our measurements on different subsets of the data, so that we may better compare our constraints to other investigations.

We check the accuracy with which our fitting methodology can recover disk parameters by creating a mock SEGUE sample from equation~\ref{eq:model}, and then analyzing it in the same way as we have analyzed the SEGUE G-dwarf sample. Our modeling methodology successfully recovers the correct thin- and thick-disk parameters within the 1$\sigma$ error bars. This exercise demonstrates that our Milky Way constraints do not contain systematic errors due to the methodology. However, there may be systematic errors in our constraints that arise from errors in the SEGUE weights we use. Although we do not expect these errors to be large given the fairly homogeneous nature of the SEGUE G-dwarf sample in the narrow distance range that we study, we cannot guarantee that these systematic errors are smaller than our statistical errors. The main point to emphasize is that the high statistical precision of our measurements (2-3\% for the scale heights and 8-20\% for the scale lengths) proves the constraining power of the correlation function statistic for Galactic studies. We note that our statistical precision is still considerably lower than that reported by \citet{Bovy:2012}, which is three to five times higher. This is most likely due to the fact that we measure the correlation function of each SEGUE line-of-sight separately, which means that the overall variation in stellar density from one sightline to another does not contribute to our model constraints. We can improve on this by measuring a single correlation function that includes cross-sightline pairs, and we leave this to a future study.

We also investigated how well the two-disk model in Equation~\ref{eq:model} explains the measured clustering of SEGUE G dwarfs. The $\chi^2$ value for our best-fit model is 2,853 for 1,772 degrees of freedom, suggesting that the model is strongly ruled out. For comparison, we tried a single exponential disk model with only two parameters. The best-fit value of $\chi^2$ in that case is 4,384 for 1,775 degrees of freedom. The two-disk model is thus strongly preferred over the single-disk model. However, even the two-disk model is excluded by our correlation function measurements.

\section{Evidence of Substructure?}
\label{s:substructure}

\begin{figure}
	\includegraphics[scale=0.3]{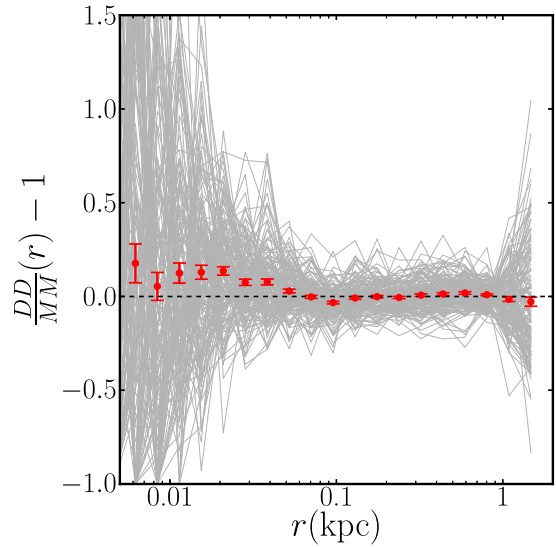}
	\caption{Correlation function residuals of SEGUE stars relative to the best-fit two-disk model. Each gray line shows the residual pair counts for one SEGUE line-of-sight. The red points show the mean residual, and error bars show the uncertainty in the mean estimated from the dispersion among the lines-of-sight. The SEGUE data clearly shows an excess clustering at small scales ($\lesssim50$~pc), suggesting possible substructures that are not included in our simple two-disk model.}
	\label{fig:compare_data_model}
\end{figure}

\begin{figure}
	\includegraphics[scale=0.32]{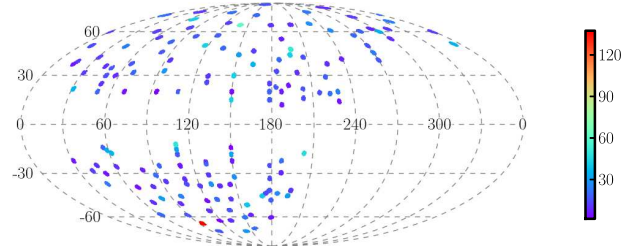}
	\caption{Sky map of $\chi^2$ values for the best-fit two-disk model. The color of each SEGUE field indicates the contribution to the global $\chi^2$ coming from that particular line-of-sight. The map reveals no obvious correlation between the goodness of fit and positions on the sky, indicating that the excess signal in the correlation function is probably not caused by field-dependent structures.}
	\label{fig:chi2_map}
\end{figure}

We next investigate the residual clustering of SEGUE stars relative to our best-fit two-disk model to see where the model fails. Figure~\ref{fig:compare_data_model} shows $\xi'(r)$ for the best-fit model along all the lines-of-sight (gray lines), as well as the mean residuals averaged over all lines-of-sight (red points). It is clear that, although our best-fit model cancels the correlation function on most scales, there remains significant excess clustering in the SEGUE data on small scales ($\lesssim50$~pc) that cannot be explained by the model. This discrepancy could be due to a number of reasons. It is possible that a smooth model of the density structure of the Milky Way can in fact fully account for our clustering measurements, but we have just adopted the wrong model. The ``correct" model could be a two-disk model with a different functional form than Equation~\ref{eq:model}. Alternatively, we may be missing one or more components, such as a third disk or, more likely, a smooth sequence of disks for stars of different ages, as suggested by \citet{Bovy:2012}. Subtle changes to the smooth density model can cause strong deviations in the correlation function, as demonstrated in Figure~\ref{fig:test_gradient}. Conversely, the excess clustering that we find could be evidence of substructure in the SEGUE data that cannot be explained by any smooth density model. For example, this signal could be due to some stars living in clusters, or could be due to the presence of large localized structures such as stellar streams. If the excess clustering is produced by localized structures on the sky, we would expect that those specific SEGUE lines-of-sight are solely responsible for the failure of our two-disk model to fit the data. We investigate this possibility in Figure~\ref{fig:chi2_map}, which displays the map of $\chi^2$ values contributed by each SEGUE field across the sky. The map does not reveal any significant spatial structure in the $\chi^2$ distribution, suggesting that the remaining signal is probably not caused by large localized structures such as stellar streams. There is one specific SEGUE field that has an abnormally high value of $\chi^2$: the red field at a Galactic latitude of $\approx-65\degree$ in Figure~\ref{fig:chi2_map}. However, removing this line-of-sight does not resolve the discrepancy between data and model.

\section{Summary and Discussion}
\label{s:summary}

In this paper we explore applying a traditional clustering statistic, the spatial two-point correlation function, to stars in the Milky Way as a probe of Galactic structure. Tests with mock samples have shown that the shape of the correlation function is sensitive to both the stellar-density gradients in the Galaxy disk and the survey geometry. We have measured the correlation function of SDSS SEGUE G-dwarf stars, which is a large and homogenous sample with well-understood selection criteria, geometry, and distance errors. By comparing our measurements to a two-disk Galactic model, our measured correlation functions yield tight constraints on the structure of the thin and thick disk of the Milky Way. Specifically, the thin- and thick-disk scale heights are determined with a precision of 3\% and 2\%, respectively, while the thin- and thick-disk scale lengths are determined with a precision of 20\% and 8\%, respectively. This high precision is achieved with spatial information alone, and it proves the strong constraining power of the correlation function. Furthermore, we have studied the residuals of the SEGUE clustering relative to our best-fit two-disk model, and have found a small but significant excess of clustering on scales less than 50~pc in the SEGUE data relative to the smooth model. This clustering may be due to imperfections in the smooth model or it may be due to the presence of substructure in the SEGUE data that cannot be described by a smooth model. The main source of systematic error in this analysis comes from uncertainties in the weights (calculated by \citealt{Schlesinger:2012}) that we use to account for sample incompleteness. Although we do not expect these uncertainties to be large, further work is needed to assess the extent to which they affect our model constraints.

There are several avenues for future work. First, the methodology we have used can be explored further and improved. For example, we can study the covariances between different data points and include them in the analysis. We can also probe larger scales by measuring pairs across neighboring lines-of-sight, instead of sticking to within one SEGUE field at a time. This should significantly improve the constraining power of the correlation function and it may detect the signatures of large structures such as stellar streams. Secondly, we can study subsamples of SEGUE stars, such as samples in specific metallicity ranges, in order to better compare our constraints against other works. We can also explore variants of the spatial correlation function, such as a metallicity- or age-weighted correlation function or a phase-space correlation function. Finally, we can further explore the cause of the discrepancy between the clustering of SEGUE stars and the two-disk model by exploring a larger family of smooth Galactic models.

\acknowledgments 
We thank Jonathan Bird, Jo Bovy, Jennifer Piscionere, Manodeep Sinha, and David Weinberg for useful discussions and suggestions.
Q.M. and A.A.B. acknowledge support from a Vanderbilt University Discovery grant. We also acknowledge the Vanderbilt Advanced Computing Center for Research and Education (ACCRE) that provided some of the computational resources used in this study. T.C.B. acknowledges partial support for this work from PHY 08-22648; Physics Frontier Center/Joint Institute for Nuclear Astrophysics (JINA) and PHY 14-30152; Physics Frontier Center/JINA Center for the Evolution of the Elements (JINA-CEE), awarded by the US National Science Foundation.

Funding for SDSS-III has been provided by the Alfred~P.~Sloan Foundation, the Participating Institutions, the National Science Foundation, and the U.S. Department of Energy Office of Science. The SDSS-III web site is http://www.sdss3.org/. 

SDSS-III is managed by the Astrophysical Research Consortium for the Participating Institutions of the SDSS-III Collaboration including the University of Arizona, the Brazilian Participation Group, Brookhaven National Laboratory, University of Cambridge, Carnegie Mellon University, University of Florida, the French Participation Group, the German Participation Group, Harvard University, the Instituto de Astrofisica de Canarias, the Michigan State/Notre Dame/JINA Participation Group, Johns Hopkins University, Lawrence Berkeley National Laboratory, Max Planck Institute for Astrophysics, Max Planck Institute for Extraterrestrial Physics, New Mexico State University, New York University, Ohio State University, Pennsylvania State University, University of Portsmouth, Princeton University, the Spanish Participation Group, University of Tokyo, University of Utah, Vanderbilt University, University of Virginia, University of Washington, and Yale University.

\bibliographystyle{apj}
\bibliography{starclustering}

\begin{thebibliography}{30}
\expandafter\ifx\csname natexlab\endcsname\relax\def\natexlab#1{#1}\fi

\bibitem[{{An} {et~al.}(2009){An}, {Pinsonneault}, {Masseron}, {Delahaye},
  {Johnson}, {Terndrup}, {Beers}, {Ivans}, \& {Ivezi{\'c}}}]{An:2009}
{An}, D., {et~al.} 2009, \apj, 700, 523

\bibitem[{{Anderson} {et~al.}(2014){Anderson}, {Aubourg}, {Bailey}, {Beutler},
  {Bhardwaj}, {Blanton}, {Bolton}, {Brinkmann}, {Brownstein}, {Burden},
  {Chuang}, {Cuesta}, {Dawson}, {Eisenstein}, {Escoffier}, {Gunn}, {Guo}, {Ho},
  {Honscheid}, {Howlett}, {Kirkby}, {Lupton}, {Manera}, {Maraston}, {McBride},
  {Mena}, {Montesano}, {Nichol}, {Nuza}, {Olmstead}, {Padmanabhan},
  {Palanque-Delabrouille}, {Parejko}, {Percival}, {Petitjean}, {Prada},
  {Price-Whelan}, {Reid}, {Roe}, {Ross}, {Ross}, {Sabiu}, {Saito}, {Samushia},
  {S{\'a}nchez}, {Schlegel}, {Schneider}, {Scoccola}, {Seo}, {Skibba},
  {Strauss}, {Swanson}, {Thomas}, {Tinker}, {Tojeiro}, {Maga{\~n}a}, {Verde},
  {Wake}, {Weaver}, {Weinberg}, {White}, {Xu}, {Y{\`e}che}, {Zehavi}, \&
  {Zhao}}]{Anderson:2014}
{Anderson}, L., {et~al.} 2014, \mnras, 441, 24

\bibitem[{{Belokurov} {et~al.}(2006){Belokurov}, {Zucker}, {Evans}, {Gilmore},
  {Vidrih}, {Bramich}, {Newberg}, {Wyse}, {Irwin}, {Fellhauer}, {Hewett},
  {Walton}, {Wilkinson}, {Cole}, {Yanny}, {Rockosi}, {Beers}, {Bell},
  {Brinkmann}, {Ivezi{\'c}}, \& {Lupton}}]{Belokurov:2006}
{Belokurov}, V., {et~al.} 2006, \apjl, 642, L137

\bibitem[{{Bensby} {et~al.}(2011){Bensby}, {Alves-Brito}, {Oey}, {Yong}, \&
  {Mel{\'e}ndez}}]{Bensby:2011}
{Bensby}, T., {Alves-Brito}, A., {Oey}, M.~S., {Yong}, D., \& {Mel{\'e}ndez},
  J. 2011, \apjl, 735, L46

\bibitem[{{Bovy} {et~al.}(2012{\natexlab{a}}){Bovy}, {Rix}, \&
  {Hogg}}]{Bovy:2012a}
{Bovy}, J., {Rix}, H.-W., \& {Hogg}, D.~W. 2012{\natexlab{a}}, \apj, 751, 131

\bibitem[{{Bovy} {et~al.}(2012{\natexlab{b}}){Bovy}, {Rix}, {Liu}, {Hogg},
  {Beers}, \& {Lee}}]{Bovy:2012}
{Bovy}, J., {Rix}, H.-W., {Liu}, C., {Hogg}, D.~W., {Beers}, T.~C., \& {Lee},
  Y.~S. 2012{\natexlab{b}}, \apj, 753, 148

\bibitem[{{Carollo} {et~al.}(2010){Carollo}, {Beers}, {Chiba}, {Norris},
  {Freeman}, {Lee}, {Ivezi{\'c}}, {Rockosi}, \& {Yanny}}]{Carollo:2010}
{Carollo}, D., {et~al.} 2010, \apj, 712, 692

\bibitem[{{Cheng} {et~al.}(2012){Cheng}, {Rockosi}, {Morrison}, {Lee}, {Beers},
  {Bizyaev}, {Harding}, {Malanushenko}, {Malanushenko}, {Oravetz}, {Pan},
  {Schlesinger}, {Schneider}, {Simmons}, \& {Weaver}}]{Cheng:2012}
{Cheng}, J.~Y., {et~al.} 2012, \apj, 752, 51

\bibitem[{{Cooper} {et~al.}(2011){Cooper}, {Cole}, {Frenk}, \&
  {Helmi}}]{Cooper:2011}
{Cooper}, A.~P., {Cole}, S., {Frenk}, C.~S., \& {Helmi}, A. 2011, \mnras, 417,
  2206

\bibitem[{{de Jong} {et~al.}(2010){de Jong}, {Yanny}, {Rix}, {Dolphin},
  {Martin}, \& {Beers}}]{deJong:2010}
{de Jong}, J.~T.~A., {Yanny}, B., {Rix}, H.-W., {Dolphin}, A.~E., {Martin},
  N.~F., \& {Beers}, T.~C. 2010, \apj, 714, 663

\bibitem[{{De Propris} {et~al.}(2010){De Propris}, {Harrison}, \&
  {Mares}}]{DePropris:2010}
{De Propris}, R., {Harrison}, C.~D., \& {Mares}, P.~J. 2010, \apj, 719, 1582

\bibitem[{{Doinidis} \& {Beers}(1989)}]{Doinidis:1989}
{Doinidis}, S.~P., \& {Beers}, T.~C. 1989, \apjl, 340, L57

\bibitem[{{Eisenstein} {et~al.}(2011){Eisenstein}, {Weinberg}, {Agol},
  {Aihara}, {Allende Prieto}, {Anderson}, {Arns}, {Aubourg}, {Bailey},
  {Balbinot}, \& et~al.}]{Eisenstein:2011}
{Eisenstein}, D.~J., {et~al.} 2011, \aj, 142, 72

\bibitem[{{Gunn} {et~al.}(2006){Gunn}, {Siegmund}, {Mannery}, {Owen}, {Hull},
  {Leger}, {Carey}, {Knapp}, {York}, {Boroski}, {Kent}, {Lupton}, {Rockosi},
  {Evans}, {Waddell}, {Anderson}, {Annis}, {Barentine}, {Bartoszek}, {Bastian},
  {Bracker}, {Brewington}, {Briegel}, {Brinkmann}, {Brown}, {Carr},
  {Czarapata}, {Drennan}, {Dombeck}, {Federwitz}, {Gillespie}, {Gonzales},
  {Hansen}, {Harvanek}, {Hayes}, {Jordan}, {Kinney}, {Klaene}, {Kleinman},
  {Kron}, {Kresinski}, {Lee}, {Limmongkol}, {Lindenmeyer}, {Long}, {Loomis},
  {McGehee}, {Mantsch}, {Neilsen}, {Neswold}, {Newman}, {Nitta}, {Peoples},
  {Pier}, {Prieto}, {Prosapio}, {Rivetta}, {Schneider}, {Snedden}, \&
  {Wang}}]{Gunn:2006}
{Gunn}, J.~E., {et~al.} 2006, \aj, 131, 2332

\bibitem[{{Ivezi{\'c}} {et~al.}(2012){Ivezi{\'c}}, {Beers}, \&
  {Juri{\'c}}}]{Ivezic:2012}
{Ivezi{\'c}}, {\v Z}., {Beers}, T.~C., \& {Juri{\'c}}, M. 2012, \araa, 50, 251

\bibitem[{{Juri{\'c}} {et~al.}(2008){Juri{\'c}}, {Ivezi{\'c}}, {Brooks},
  {Lupton}, {Schlegel}, {Finkbeiner}, {Padmanabhan}, {Bond}, {Sesar},
  {Rockosi}, {Knapp}, {Gunn}, {Sumi}, {Schneider}, {Barentine}, {Brewington},
  {Brinkmann}, {Fukugita}, {Harvanek}, {Kleinman}, {Krzesinski}, {Long},
  {Neilsen}, {Nitta}, {Snedden}, \& {York}}]{Juric:2008}
{Juri{\'c}}, M., {et~al.} 2008, \apj, 673, 864

\bibitem[{{Kordopatis} {et~al.}(2013){Kordopatis}, {Gilmore}, {Steinmetz},
  {Boeche}, {Seabroke}, {Siebert}, {Zwitter}, {Binney}, {de Laverny},
  {Recio-Blanco}, {Williams}, {Piffl}, {Enke}, {Roeser}, {Bijaoui}, {Wyse},
  {Freeman}, {Munari}, {Carrillo}, {Anguiano}, {Burton}, {Campbell}, {Cass},
  {Fiegert}, {Hartley}, {Parker}, {Reid}, {Ritter}, {Russell}, {Stupar},
  {Watson}, {Bienaym{\'e}}, {Bland-Hawthorn}, {Gerhard}, {Gibson}, {Grebel},
  {Helmi}, {Navarro}, {Conrad}, {Famaey}, {Faure}, {Just}, {Kos}, {Matijevi{\v
  c}}, {McMillan}, {Minchev}, {Scholz}, {Sharma}, {Siviero}, {de Boer}, \& {{\v
  Z}erjal}}]{Kordopatis:2013}
{Kordopatis}, G., {et~al.} 2013, \aj, 146, 134

\bibitem[{{Longhitano} \& {Binggeli}(2010)}]{Longhitano:2010}
{Longhitano}, M., \& {Binggeli}, B. 2010, \aap, 509, A46

\bibitem[{{Lopez-Corredoira} {et~al.}(1998){Lopez-Corredoira}, {Garzon},
  {Hammersley}, \& {Mahoney}}]{Lopez-Corredoira:1998}
{Lopez-Corredoira}, M., {Garzon}, F., {Hammersley}, P.~L., \& {Mahoney}, T.~J.
  1998, \mnras, 301, 289

\bibitem[{{Morrison} {et~al.}(2000){Morrison}, {Mateo}, {Olszewski}, {Harding},
  {Dohm-Palmer}, {Freeman}, {Norris}, \& {Morita}}]{Morrison:2000}
{Morrison}, H.~L., {Mateo}, M., {Olszewski}, E.~W., {Harding}, P.,
  {Dohm-Palmer}, R.~C., {Freeman}, K.~C., {Norris}, J.~E., \& {Morita}, M.
  2000, \aj, 119, 2254

\bibitem[{{Peebles}(1973)}]{Peebles:1973}
{Peebles}, P.~J.~E. 1973, \apj, 185, 413

\bibitem[{{Rix} \& {Bovy}(2013)}]{Rix:2013}
{Rix}, H.-W., \& {Bovy}, J. 2013, \aapr, 21, 61

\bibitem[{{Schlegel} {et~al.}(1998){Schlegel}, {Finkbeiner}, \&
  {Davis}}]{Schlegel:1998}
{Schlegel}, D.~J., {Finkbeiner}, D.~P., \& {Davis}, M. 1998, \apj, 500, 525

\bibitem[{{Schlesinger} {et~al.}(2012){Schlesinger}, {Johnson}, {Rockosi},
  {Lee}, {Morrison}, {Sch{\"o}nrich}, {Allende Prieto}, {Beers}, {Yanny},
  {Harding}, {Schneider}, {Chiappini}, {da Costa}, {Maia}, {Minchev},
  {Rocha-Pinto}, \& {Santiago}}]{Schlesinger:2012}
{Schlesinger}, K.~J., {et~al.} 2012, \apj, 761, 160

\bibitem[{{Skrutskie} {et~al.}(2006){Skrutskie}, {Cutri}, {Stiening},
  {Weinberg}, {Schneider}, {Carpenter}, {Beichman}, {Capps}, {Chester},
  {Elias}, {Huchra}, {Liebert}, {Lonsdale}, {Monet}, {Price}, {Seitzer},
  {Jarrett}, {Kirkpatrick}, {Gizis}, {Howard}, {Evans}, {Fowler}, {Fullmer},
  {Hurt}, {Light}, {Kopan}, {Marsh}, {McCallon}, {Tam}, {Van Dyk}, \&
  {Wheelock}}]{Skrutskie:2006}
{Skrutskie}, M.~F., {et~al.} 2006, \aj, 131, 1163

\bibitem[{{Smee} {et~al.}(2013){Smee}, {Gunn}, {Uomoto}, {Roe}, {Schlegel},
  {Rockosi}, {Carr}, {Leger}, {Dawson}, {Olmstead}, {Brinkmann}, {Owen},
  {Barkhouser}, {Honscheid}, {Harding}, {Long}, {Lupton}, {Loomis}, {Anderson},
  {Annis}, {Bernardi}, {Bhardwaj}, {Bizyaev}, {Bolton}, {Brewington}, {Briggs},
  {Burles}, {Burns}, {Castander}, {Connolly}, {Davenport}, {Ebelke}, {Epps},
  {Feldman}, {Friedman}, {Frieman}, {Heckman}, {Hull}, {Knapp}, {Lawrence},
  {Loveday}, {Mannery}, {Malanushenko}, {Malanushenko}, {Merrelli}, {Muna},
  {Newman}, {Nichol}, {Oravetz}, {Pan}, {Pope}, {Ricketts}, {Shelden},
  {Sandford}, {Siegmund}, {Simmons}, {Smith}, {Snedden}, {Schneider},
  {SubbaRao}, {Tremonti}, {Waddell}, \& {York}}]{Smee:2013}
{Smee}, S.~A., {et~al.} 2013, \aj, 146, 32

\bibitem[{{Starkenburg} {et~al.}(2009){Starkenburg}, {Helmi}, {Morrison},
  {Harding}, {van Woerden}, {Mateo}, {Olszewski}, {Sivarani}, {Norris},
  {Freeman}, {Shectman}, {Dohm-Palmer}, {Frey}, \&
  {Oravetz}}]{Starkenburg:2009}
{Starkenburg}, E., {et~al.} 2009, \apj, 698, 567

\bibitem[{{Xue} {et~al.}(2011){Xue}, {Rix}, {Yanny}, {Beers}, {Bell}, {Zhao},
  {Bullock}, {Johnston}, {Morrison}, {Rockosi}, {Koposov}, {Kang}, {Liu},
  {Luo}, {Lee}, \& {Weaver}}]{Xue:2011}
{Xue}, X.-X., {et~al.} 2011, \apj, 738, 79

\bibitem[{{Yanny} {et~al.}(2009){Yanny}, {Rockosi}, {Newberg}, {Knapp},
  {Adelman-McCarthy}, {Alcorn}, {Allam}, {Allende Prieto}, {An}, {Anderson},
  {Anderson}, {Bailer-Jones}, {Bastian}, {Beers}, {Bell}, {Belokurov},
  {Bizyaev}, {Blythe}, {Bochanski}, {Boroski}, {Brinchmann}, {Brinkmann},
  {Brewington}, {Carey}, {Cudworth}, {Evans}, {Evans}, {Gates}, {G{\"a}nsicke},
  {Gillespie}, {Gilmore}, {Nebot Gomez-Moran}, {Grebel}, {Greenwell}, {Gunn},
  {Jordan}, {Jordan}, {Harding}, {Harris}, {Hendry}, {Holder}, {Ivans},
  {Ivezi{\v c}}, {Jester}, {Johnson}, {Kent}, {Kleinman}, {Kniazev},
  {Krzesinski}, {Kron}, {Kuropatkin}, {Lebedeva}, {Lee}, {French Leger},
  {L{\'e}pine}, {Levine}, {Lin}, {Long}, {Loomis}, {Lupton}, {Malanushenko},
  {Malanushenko}, {Margon}, {Martinez-Delgado}, {McGehee}, {Monet}, {Morrison},
  {Munn}, {Neilsen}, {Nitta}, {Norris}, {Oravetz}, {Owen}, {Padmanabhan},
  {Pan}, {Peterson}, {Pier}, {Platson}, {Re Fiorentin}, {Richards}, {Rix},
  {Schlegel}, {Schneider}, {Schreiber}, {Schwope}, {Sibley}, {Simmons},
  {Snedden}, {Allyn Smith}, {Stark}, {Stauffer}, {Steinmetz}, {Stoughton},
  {SubbaRao}, {Szalay}, {Szkody}, {Thakar}, {Sivarani}, {Tucker}, {Uomoto},
  {Vanden Berk}, {Vidrih}, {Wadadekar}, {Watters}, {Wilhelm}, {Wyse}, {Yarger},
  \& {Zucker}}]{Yanny:2009}
{Yanny}, B., {et~al.} 2009, \aj, 137, 4377

\bibitem[{{York} {et~al.}(2000){York}, {Adelman}, {Anderson}, {Anderson},
  {Annis}, {Bahcall}, {Bakken}, {Barkhouser}, {Bastian}, {Berman}, {Boroski},
  {Bracker}, {Briegel}, {Briggs}, {Brinkmann}, {Brunner}, {Burles}, {Carey},
  {Carr}, {Castander}, {Chen}, {Colestock}, {Connolly}, {Crocker}, {Csabai},
  {Czarapata}, {Davis}, {Doi}, {Dombeck}, {Eisenstein}, {Ellman}, {Elms},
  {Evans}, {Fan}, {Federwitz}, {Fiscelli}, {Friedman}, {Frieman}, {Fukugita},
  {Gillespie}, {Gunn}, {Gurbani}, {de Haas}, {Haldeman}, {Harris}, {Hayes},
  {Heckman}, {Hennessy}, {Hindsley}, {Holm}, {Holmgren}, {Huang}, {Hull},
  {Husby}, {Ichikawa}, {Ichikawa}, {Ivezi{\'c}}, {Kent}, {Kim}, {Kinney},
  {Klaene}, {Kleinman}, {Kleinman}, {Knapp}, {Korienek}, {Kron}, {Kunszt},
  {Lamb}, {Lee}, {Leger}, {Limmongkol}, {Lindenmeyer}, {Long}, {Loomis},
  {Loveday}, {Lucinio}, {Lupton}, {MacKinnon}, {Mannery}, {Mantsch}, {Margon},
  {McGehee}, {McKay}, {Meiksin}, {Merelli}, {Monet}, {Munn}, {Narayanan},
  {Nash}, {Neilsen}, {Neswold}, {Newberg}, {Nichol}, {Nicinski}, {Nonino},
  {Okada}, {Okamura}, {Ostriker}, {Owen}, {Pauls}, {Peoples}, {Peterson},
  {Petravick}, {Pier}, {Pope}, {Pordes}, {Prosapio}, {Rechenmacher}, {Quinn},
  {Richards}, {Richmond}, {Rivetta}, {Rockosi}, {Ruthmansdorfer}, {Sandford},
  {Schlegel}, {Schneider}, {Sekiguchi}, {Sergey}, {Shimasaku}, {Siegmund},
  {Smee}, {Smith}, {Snedden}, {Stone}, {Stoughton}, {Strauss}, {Stubbs},
  {SubbaRao}, {Szalay}, {Szapudi}, {Szokoly}, {Thakar}, {Tremonti}, {Tucker},
  {Uomoto}, {Vanden Berk}, {Vogeley}, {Waddell}, {Wang}, {Watanabe},
  {Weinberg}, {Yanny}, {Yasuda}, \& {SDSS Collaboration}}]{York:2000}
{York}, D.~G., {et~al.} 2000, \aj, 120, 1579

\end{thebibliography}

\end{document}